\documentclass[12pt]{article}

\usepackage[utf8]{inputenc}
\usepackage{charter}
\usepackage{fullpage}
\usepackage{hyperref}
\usepackage{graphicx}
\usepackage{lipsum}
\usepackage[caption=false]{subfig}
\usepackage[sort&compress,numbers]{natbib}
\usepackage{hyperref}
\hypersetup{hidelinks}
\usepackage[total={6.5in,9in},top=1in,headsep=0.1in,headheight=1in]{geometry}
\usepackage[dvipsnames]{xcolor}
\usepackage{amsmath}
\usepackage{amssymb}
\usepackage{aas_macros}
\usepackage{siunitx}

\newcommand\snowmass{
\begin{center}
  \rule[-0.2in]{\hsize}{0.01in}\\
  \rule{\hsize}{0.01in}\\
  \vskip 0.1in
  Submitted to the Proceedings of the US Community Study\\ 
  on the Future of Particle Physics (Snowmass 2021)\\
  \rule{\hsize}{0.01in}\\
  \rule[+0.2in]{\hsize}{0.01in}\\[-2em]
\end{center}
}

\usepackage[firstpage=true]{background}
\backgroundsetup{contents={\parbox{6.5in}{\snowmass}}, scale=1,placement=top,opacity=1,color=black,position={3.25in,1.2in}}

\usepackage{fancyhdr}
\fancypagestyle{plain}{%
  \fancyhf{}%
  \fancyhead[C]{}
  \fancyfoot[C]{\thepage}
}

\fancypagestyle{empty}{%
  \fancyhf{}%
  \fancyhead[C]{{\it Probing dark matter with small-scale astrophysical observations}}
  \fancyfoot[C]{\thepage}
}
\title{Snowmass2021 Cosmic Frontier White Paper:
\\Probing dark matter with small-scale astrophysical observations}
\date{}

\usepackage{authblk}

\newcommand{\Neff}{\ensuremath{N_\mathrm{eff}}}

\author{Richard Brito}
\affil{CENTRA, Departamento de F\'{\i}sica, Instituto Superior T\'ecnico -- IST, Universidade de Lisboa -- UL, Avenida Rovisco Pais 1, 1049 Lisboa, Portugal}

\author{Sukanya Chakrabarti}
\affil{School of Physics and Astronomy, Rochester Institute of Technology, Rochester, NY 14623 USA}

\author{S\'{e}bastien Clesse}
\affil{Service de Physique Th\'eorique, Universit\'e Libre de Bruxelles, Boulevard du Triomphe CP225, B-10503 Brussels, Belgium}

\author{Cora Dvorkin}
\affil{Department of Physics, Harvard University, 17 Oxford Street, Cambridge, MA 02138, USA}

\author{Juan Garc\'{i}a-Bellido}
\affil{Instituto de F\'isica Te\'orica UAM/CSIC, Universidad Aut\'onoma de Madrid, 28049 Madrid, Spain}

\author{Joel Meyers}
\affil{Department of Physics, Southern Methodist University, Dallas, TX 75275, USA}

\author{Ken K. Y. Ng}
\affil{LIGO Lab, Massachusetts Institute of Technology, Cambridge, Massachusetts 02139, USA}
\affil{Kavli Institute for Astrophysics and Space Research, Department of Physics, Massachusetts Institute of Technology, Cambridge, Massachusetts 02139, USA}

\author{Andrew L. Miller}
\affil{Universit\'e catholique de Louvain, B-1348 Louvain-la-Neuve, Belgium}

\author{Sarah Shandera}
\affil{Institute for Gravitation and the Cosmos, The Pennsylvania State University, University Park, PA 16802, USA}
\affil{Department of Physics, The Pennsylvania State University, University Park, PA, 16802, USA}

\author{Ling Sun}
\affil{OzGrav-ANU, Centre for Gravitational Astrophysics, College of Science,The Australian National University, Australian Capital Territory 2601, Australia}

\begin{document}

\maketitle

\begin{abstract}
The current understanding of dark matter comes largely from measurements of the total matter content in the universe, from the distribution of gravitating matter on very large scales, and from rotation curves and velocity dispersions on sub-galactic scales. However, small-scale structure may well hold the key to unlocking the particle physics of the still-mysterious 85\% of matter in the universe. Novel small-scale astrophysical probes of new particles and dark matter will become possible with a large data volume of high-precision measurements enabled by next-generation gravitational-wave detectors and advanced astrometry instruments. Cosmic microwave background and large-scale structure surveys will provide complementary constraints on dark matter models with unique small-scale signatures. We lay out the studies of small-scale structures and compact objects as dark matter probes, and summarize the requirements to achieve the goals.
\end{abstract}

\tableofcontents

\section{Executive summary}
The last decade brought the first successful detection of gravitational waves (GWs) \cite{LIGOScientific:2016aoc} and a dramatic improvement in high-precision astrometry \cite{Gaia:2016zol}. As GW detectors mature and new instruments to measure proper motions and spectroscopy are deployed in the next few years, they will provide the first high data-volume probes of the ultra small-scale structure of the universe. This data will enable searches for novel populations of compact objects that cannot be formed in stellar processes, will harness the extreme gravitational environment around black holes to constrain new ultralight particles, and will provide precise determinations of the dark matter distribution down to earth-mass scales. Coordination of the communities that study compact objects and astrometry with those that study dark matter is required in order to realize the science potential of this data. Here we lay out the case for small-scale structure and compact-object data as dark matter probes, covering a few key questions about the nature of the Universe:
(1) Is dark matter made of compact objects such as sub-solar mass black holes and of primordial origin? 
(2) Is the dark matter composed of interacting or dissipative beyond-Standard-Model particles? 
(3) What are the GW signatures of dark matter?
These perspectives should inform future decisions about the design and science reach of surveys of ultra small-scale structure, including compact objects. 

Our current understanding of dark matter comes from measurements of the total matter content in the universe, from the distribution of gravitating matter on very large scales where baryonic feedback is relatively unimportant, and from rotation curves and velocity dispersions on sub-galactic scales. There are hints from some of these measurements that dark matter may not fit traditional models \cite{Bullock:2017xww}, but the complex nature of baryonic feedback makes definitive statements elusive. However, the properties of structure on smaller scales are predicted to vary even more dramatically between dark matter scenarios. Small-scale structure may well hold the key to unlocking the early universe physics of the still-mysterious 85\% of matter in the universe.

\begin{figure}[tbh!]
    \centering
    \includegraphics[width=\textwidth]{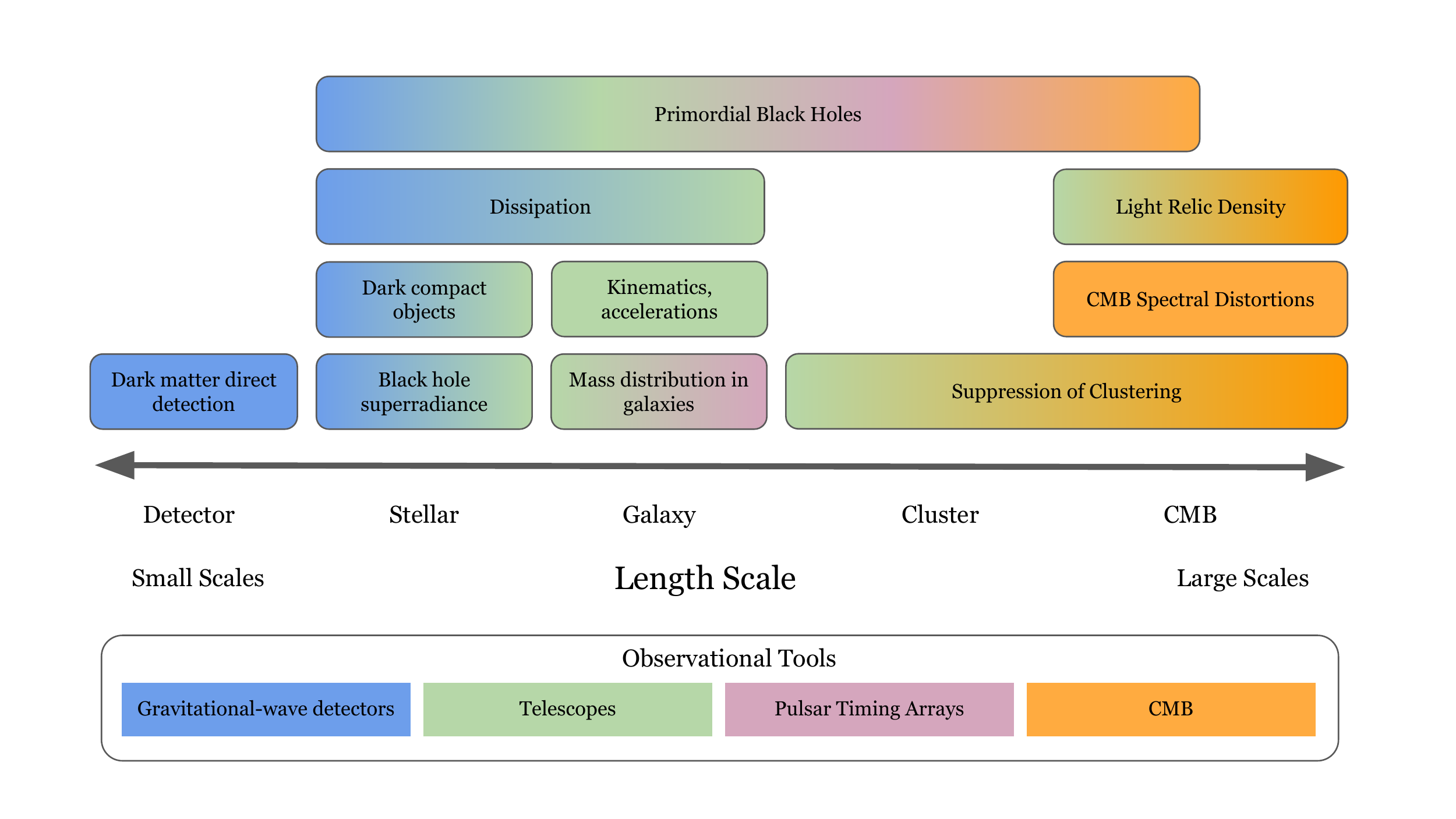}
    \caption{The diagram shows phenomena probing dark  matter, on length scales ranging from terrestrial detectors to the cosmic microwave background (CMB). The colors indicate the new generation of observational tools that can enable the cosmic probes of these phenomena. The unprecedented measurements on sub-galactic small scales, combined with complementary advances on large scales, will lead to strong constraints and largely improved understanding of dark matter.  Future telescope facilities, including TMT, GMT, MegaMapper, MSE, and SpecTel, will be sensitive to the dark matter halo mass function down to $\sim 10^{6}~M_{\odot}$ \citep{Chakrabartietal2022}. }
    \label{fig:wp5}
\end{figure}

The data from the future observing runs of the LIGO-Virgo-KAGRA GW detector network~\cite{LIGO2014,Virgo2014,KAGRA:2013rdx}, from the Rubin Observatory Legacy Survey of Space and Time (LSST)~\cite{RubinLSST}, and the Nancy Grace Roman Space Telescope~\cite{Spergel:2015sza}, will lead to strong constraints on dark matter from small-scale structure. A careful understanding of the implications of that data will be critical in planning for and developing the next stage of instruments, including the next-generation ground-based GW detectors like Cosmic Explorer \cite{Evans2021horizon} and Einstein Telescope \cite{Maggiore:1999vm}, space-based detectors including Laser Interferometer Space Antenna (LISA) \cite{2013arXiv1305.5720E}, TianQin \cite{TianQin}, and the DECi-hertz Interferometer Gravitational wave Observatory (DECIGO) \cite{Kawamura_2008}, and a possible successor to Gaia~\cite{Gaia:2016zol} (see more details about future facilities in Refs.~\cite{snowmass_facility,snowmass_GWfacility}).
The novel searches on various length scales, together with the observational enablers, are laid out in Figure~\ref{fig:wp5}. 

Recent work has also shown significant advances in constraining dark objects at the stellar mass scales.  Large-scale spectroscopic surveys, such as the Apache Point Observatory Galactic Evolution Experiment (APOGEE), have been analyzed to search for stellar mass black holes \citep{Thompson2019,Jayasinghe2021,Jayasinghe2022}. Microlensing survey programs, e.g., the Optical Gravitational Lensing Experiment (OGLE) and Microlensing Observations in Astrophysics (MOA), have been analyzed to identify long-duration microlensing events that have been attributed as arising due to black holes \cite{WyrzMandel2020}.  Additionally, Hubble Space Telescope (HST) follow-up of a source identified independently in OGLE and MOA provided a measurement of the small astrometric shift ($\sim$ milliarcseconds), which enables a measurement of the mass of lens that is not degenerate with the transverse velocity \citep{Sahu2022}.  In external galaxies, quasar microlensing provides a route to probe mass distributions at the planet-mass scale  \citep{Dai2018,Bhatiani2019}.  Advances in technology should also enable constraints on dark matter sub-structure by measuring Galactic accelerations directly, using pulsar timing \citep{Chakrabarti2021}, extreme-precision radial velocity observations \citep{Chakrabarti2020}, and by measuring the small shift in the eclipse time induced by the Galactic potential \citep{Chakrabarti2021}.  These kind of time-series precision measurements of accelerations can be carried out across the Galaxy (and possibly beyond) with the Extremely Large Telescopes (ELTs)---the Thirty Meter Telescope (TMT) and the Giant Magellan Telescope (GMT).  These facilities as well as wide-field spectroscopic facilities, including the Dark Energy Spectroscopic Instrument DESI-II, MegaMapper, Maunakea Spectroscopic Explorer (MSE), and SpecTel, will enable constraints on dark matter sub-structure down to $\sim 10^{6}~M_{\odot}$ using future facilities \citep{Chakrabartietal2022}.  Future Cosmic Microwave Background (CMB) facilities, such as CMB-S4~\citep{Snowmass2021:CMBS4} and CMB-HD~\citep{CMB-HD:2022bsz}, will also be able to measure the matter power spectrum
on small scales using gravitational lensing of the microwave background radiation \citep{Chakrabartietal2022}.

In addition to advances in the electromagnetic observations of structure, the direct detection of GWs by the LIGO, Virgo, and KAGRA collaborations provides a completely new avenue to understand the matter content of the universe. GW observations provide the first data that can treat dark matter on precisely the same footing as the light-emitting matter of the Standard Model. As a result, they open a new direction in dark matter physics that does not rely on any non-gravitational coupling between dark matter and the Standard Model. Populations incompatible with standard stellar evolution, such as subsolar mass black holes, may be detected through the now-familiar observation of binary mergers. Alternatively, the extreme gravitational effects of a black hole on its environment can generate a new gravitational signal of light fields via superradiance. In addition, the GW detectors themselves double as a new kind of direct detection experiment. Also see discussions in Ref.~\cite{snowmass_fundamental}.

While the most detailed insights on non-minimal dark sectors may come from studying small-scale structure and compact objects, scenarios that introduce additional particles are also constrained by measurements of the total matter and radiation content of the universe. Next generation surveys of the CMB~\cite{Chang:2022tzj,Snowmass2021:CMBS4,CMB-HD:2022bsz} will provide stringent new limits on the radiation density ($N_{\rm eff}$)~\cite{Snowmass2021:LightRelics}. The combination of data from future galaxy surveys and CMB surveys will precisely measure the clustering of matter, providing excellent sensitivity to the presence of matter that does not cluster in the same way as ordinary cold dark matter, including massive neutrinos, warm dark matter, ultralight axion dark matter, and light but massive relics. We discuss the potential for complementarity between advances in CMB, large-scale structure surveys, and small-scale structure measurements in the coming decade.

All of these advances will become accessible with the use of multiple facilities in the whole range displayed in Fig.~\ref{fig:wp5}, from new GW detectors, described at length in the Snowmass paper~\cite{snowmass_GWfacility}, to the next generation telescopes, pulsar timing arrays~\cite{snowmass_facility} and CMB experiments~\cite{Chang:2022tzj}. All of these facilities will act in synergy as a multiprobe exploration of the physics of dark matter, testing both massive compact objects as dark matter and a variety of scenarios that introduce new particles into the hot big bang universe. 


\section{Background and motivation} 
\subsection{Probes of ultra-small-scale structure} 
Kinematic probes that allow the inference of the Galactic and dark matter structure on increasingly small scales take several forms. These include the 3d velocities and positions of stars in the Galaxy and stellar streams \citep{Sandeson2017,Lipnicky2017,Helmi2018,Bonaca2019,Chakrabarti2019,Widmark2020}, which provide constraints on fundamental Galactic parameters, the Galactic potential, the orbits of dwarf galaxies, dark matter sub-structure on small scales, as well detection of the smallest gravitationally bound objects \citep{Simon2019}, and possible collisions of dwarf galaxies \cite{Silk:2019jbt,Lee:2021wfv} that can be used to test for collisional self-interaction of dark matter \citep{Loeb2022}.  However, kinematic probes may not accurately represent the time-dependent Galactic potential.  Direct acceleration measurements using time-series extremely precise observations using multiple different techniques have recently become possible. These direct acceleration measurements provide the most direct probes into the mass distributions in galaxies \citep{Chakrabarti2020,Chakrabarti2021,ChakrabartiET}.  These techniques include pulsar timing \citep{Chakrabarti2021}, extreme-precision radial velocity observations that are now enabled with current generation high-resolution spectrographs (and others that will come online over the next several years) \citep{SilverwoodEasther2019,Chakrabarti2020}, and eclipse timing \citep{ChakrabartiET} to measure the very small accelerations of stars that live within the gravitational potential of the Milky Way (about 10 cm/s/decade) (also see Ref.~\cite{snowmass_facility}).

While galactic interiors are regimes governed by the interplay of many complex processes, some probes of    structure on the smallest scales may provide remarkably clean data about dark matter. For example, dark matter models can predict novel compact objects, including black holes, with mass and compactness for which there are no formation channels via stellar processes in the standard model. The populations and distributions of any such objects may well be affected by baryonic processes, but their existence would unambiguously be linked to new types of particles in the universe \cite{Salucci2019}. For example, GW detectors are sensitive to ultracompact objects with masses well below a solar mass, and any observed subsolar mass black holes would either of a primordial origin or formed dynamically in the late universe from particle dark matter \cite{Magee:2018opb, LIGOScientific:2018glc, LIGOScientific:2019kan, LIGOScientific:2021job, Nitz:2021vqh, Nitz:2021mzz, Nitz:2020bdb, Nitz:2022ltl}. 

While the primary objective of GW detectors is to detect astrophysical signals, they can also act as particle physics laboratories to directly search for ultralight dark matter \cite{Bertone:2018krk,Bertone:2019irm} in the mass range $\mathcal{O}(10^{-14}-10^{-11})$ eV/$c^2$. In brief, different dark matter particles would interact in various ways with standard model particles in the interferometer components (in particular, the laser light, the mirrors or the beam splitter), which would lead to a differential strain on the detector. The kind of interaction and subsequent differential strain depend on the type or model of dark matter considered. 

In addition to using GW detectors directly as dark matter detectors, cosmic probes via GW observations provide an alternative novel approach to search for ultralight, beyond-Standard-Model particles that could constitute a portion or all of dark matter.
Ultralight boson particles have been predicted to solve problems in particle physics, high-energy theory, and cosmology. Such fundamental scalar (spin 0), vector (spin 1), or tensor (spin 2) fields could form macroscopic clouds around rapidly rotating black holes through a phenomenon called ``superradiance''~\cite{Penrose1969,Press1972,Bekenstein1973,Arvanitaki2010,Brito2014,Brito2015}. Such clouds are expected to emit continuous, quasimonochromatic s that fall in the sensitive range of GW detectors~\cite{Arvanitaki2011,Yoshino2014,Yoshino2015,Arvanitaki2015,Arvanitaki2017,Baryakhtar2017,Brito2017-letter,Brito2017,Baryakhtar2017,Siemonsen2020,Brito2020}. 
Given that such cosmic probes take advantage of the universal gravitational coupling of these predicted ultralight bosons, the constraints do not rely on model-dependent predictions of the particles' couplings to the Standard Model, as long as those are small enough~\cite{Fukuda:2019ewf,Baryakhtar:2020gao}.

The astrophysical GW signals emitted by boson clouds that are expected to form around black holes through superradiance largely resemble the signals produced by direct dark matter interactions with the detectors; in both cases, we expect persistent, quasimonochromatic signals in the data collected by GW detectors.
Hence, similar methods can be used to search for both types of signals resulting from dark matter particles in the same mass range.
Analyses have been conducted using data collected by existing Advanced LIGO and Virgo detectors, yielding interesting constraints on the properties of these yet-undiscovered particles. 
The future ground-based detector network, including Cosmic Explorer and Einstein Telescope, with improved sensitivity, as well as space-based GW detectors, will provide the capability to constrain a large parameter space of the ultralight bosons, or even a direct detection.
By searching for such GW and dark matter signals, ground-based detectors (e.g., \cite{LIGO2014,Virgo2014,KAGRA,et-design,CE}), and space-based detectors (e.g., \cite{LISA,TianQin}) could probe bosons with masses in the ranges of \mbox{$10^{-14} \lesssim \mu/{\rm eV} \lesssim 10^{-11}$} and \mbox{$10^{-19}\lesssim \mu/{\rm eV} \lesssim 10^{-15}$}, respectively, which are largely inaccessible to conventional experiments \cite{Arvanitaki2015,Brito2017-letter,Brito2017,Isi2019} (but also see methods like~\cite{Mukherjee2020}). 
The constraints obtained using GW detectors are of great interest and complementary to related research in particle physics, high-energy theory, and cosmology. If there is a detection, detailed measurements of the signal morphology would provide invaluable information about the properties of the new particle, and would shed light on the fascinating connection between black holes and particle physics.

Axions are pseudo-scalar bosons that are potentially detectable not only via GW observations of the phenomenon of superradiance and directly via their couplings to light in the GW interferometers \cite{nagano2019axion}, but also through their effects on small-scale structures.
Self-attractive axions may produce numerous dark matter haloes that are denser but smaller than those predicted by the $\Lambda$CDM model~\cite{Arvanitaki:2019rax}.
The halo scale mass depends on the axion mass: $\sim 5\times10^9~M_{\odot}\left(10^{-22}~\mathrm{eV}/\mu\right)^{3/2}$, and may have a soliton structure with a soft core~(e.g., \cite{Mocz:2017wlg}) instead of the cuspy Navarro-Frenk-White profile~\cite{Navarro:1996gj}.
These soliton-typed dark matter haloes can act as gravitational lenses and perturb the propagation of the background electromagnetic (EM) radiations or GWs.
Searching for the gravitational lensing signatures in the EM surveys (magnified multiple images) or the GW observations (diffraction effect) can constrain the axion mass ranging from $\mathcal{O}(10^{-14}-10^{-7})~\mathrm{eV}/c^2$ (EM lensing), $\mathcal{O}(10^{-18}-10^{-16})~\mathrm{eV}/c^2$ (GW lensing in the ground-based detectors), and $\mathcal{O}(10^{-21}-10^{-18})~\mathrm{eV}/c^2$ (GW lensing in LISA).
See Fig.~11 in Ref.~\cite{Arvanitaki:2019rax} for other possible probes and detailed constraints in the parameter space of the axion mass and the halo scale mass density.

Models for the dark sector that predict unique signatures on ultra-small scales often involve new degrees of freedom that can be searched for with integrated measures of the matter and radiation content of the universe. Upcoming CMB surveys will greatly improve the precision with which we measure the light relic density, $N_\mathrm{eff}$, making these surveys very sensitive probes of new light degrees of freedom.  Candidates for light relics that will be constrained by upcoming CMB surveys include a thermal population of axions, light particles that mediate interactions in the dark sector, sterile neutrinos, and many more~\cite{Green:2019glg,Goldberg:1986nk,Foot:1999hm, Foot:2000vy, Abazajian:2001nj, Strumia:2006db, Ackerman:2008gi, Khlopov:2008ty,Kribs:2009fy,Alves:2009nf,Kaplan:2009de, Boyarsky:2009ix, Arvanitaki:2009fg, SpierMoreiraAlves:2010err,Cadamuro:2010cz, Kaplan:2011yj,Cline:2012is,Abazajian:2012ys, CyrRacine:2012fz, Brust:2013xpv, Weinberg:2013kea, Salvio:2013iaa, Essig:2013lka,Cline:2013pca,Fan:2013yva,Fan:2013tia,Fan:2013bea,Cyr-Racine:2013fsa,Cline:2013zca,Antipin:2015xia,Kawasaki:2015ofa, Graham:2015cka, Marsh:2015xka, Baumann:2016wac, Alexander:2016aln, Arkani-Hamed:2016rle, Chacko:2016hvu, Craig:2016lyx,Kribs:2016cew,Ko:2017uyb, Rosenberg:2017qia,Ghalsasi:2017jna,Chacko:2018vss,Ibe:2018juk,Gresham:2018anj,Essig:2018pzq,Alvarez:2019nwt,Curtin:2019lhm, Curtin:2019ngc,Winch:2020cju,Cyr-Racine:2021alc,Blinov:2021mdk,Cline:2021itd, Hippert:2021fch}.  Additionally, the influence of light species that became non-relativistic after recombination (similarly to the neutrinos of the Standard Model) leads to a suppression of matter clustering on scales smaller than their free-streaming length~\cite{Green:2021xzn,Munoz:2018ajr,DePorzio:2020wcz}.  This suppression of clustering can be seen through measurements of CMB lensing~\cite{Kaplinghat:2003bh, Lewis:2006fu}, clustering and weak lensing of galaxies~\cite{Hu:1997mj,Cooray:1999rv}, and the number density of galaxy clusters~\cite{Abazajian:2011dt,Carbone:2011by,Ichiki:2011ue}.  These measurements of the light relic density and of the clustering of matter made possible by CMB and large scale structure surveys serve as complementary probes of the physics responsible for unique signatures expected on ultra-small scales.

\subsection{Observational enablers} 

With the direct observation of GWs, existing ground-based GW detectors, including Advanced Laser Interferometer Gravitational-Wave Observatory (Advanced LIGO)~\cite{LIGO2014}, Advanced Virgo~\cite{Virgo2014}, and KAGRA~\cite{KAGRA}, have opened a new era of GW astrophysics and astronomy. The use of GW detectors to probe the dark sector has already begun, while the next-generation detectors with significantly improved sensitivity, e.g., Einstein Telescope \cite{Punturo:2010zz,Hild:2010id} and Cosmic Explorer \cite{nextGen-2017,Reitze2019cosmic,Evans2021horizon}, as well as space-based GW detectors, e.g., LISA \cite{LISA}, TianQin \cite{TianQin}, and DECIGO \cite{Kawamura_2008}, will enable the detailed study of fundamental physics and potentially lead to the breakthroughs in constraining the properties and/or detecting beyond-Standard-Model particles and dark matter. 
The development and capabilities of the future GW detectors are detailed in Ref.~\cite{snowmass_GWfacility}.

The expected Galactic acceleration, i.e., the change in the line-of-sight velocity, $\Delta RV$ over a decade, experienced by stars at $\sim$ kpc distances from the Sun is $\sim$ 10 cm/s/decade \citep{Chakrabarti2020}.  The current generation of spectrographs like ESPRESSO/NEID have nearly achieved this level of RV precision, but as they are on 8m (or smaller) telescopes, they cannot reach large distances in the Galaxy at the requisite precision.  The ELTs are expected to enable acceleration measurements \emph{across} the Galaxy with instruments of comparable precision~\cite{snowmass_facility}.  Aside from their ability to reach larger distances, the ELTs should also provide a measurement of dark matter sub-structure from acceleration measurements across many different lines of sight.

Pulsar timing observations were recently analyzed by \cite{Chakrabarti2021} to directly obtain an acceleration using the time-rate of change of the binary period for a set of fourteen pulsars that were timed sufficiently precisely such that the Galactic acceleration could be measured.  These accelerations enabled a measurement of the total mid-plane density (also known as the Oort limit), the local dark matter density, and the shape of the Galactic potential traced by the pulsars.  Although these observations provided the first direct measurement of Galactic accelerations, they are limited to about $\sim$ kpc distances from the Sun.  The acceleration of the solar system has also recently been measured from \textit{Gaia} astrometry \citep{Klioner2021}. Future data releases from pulsar timing facilities, as well as future radio facilities like the Next Generation Very Large Array (ngVLA) and the Deep Synoptic Array (DSA-2000) are expected to increase the dynamic range yield measurements out to the Magellanic Clouds, as well as provide constraints on dark matter sub-structure down to the $\sim 10^{6}~M_{\odot}$ scale~\cite{snowmass_facility}.

Upcoming surveys of the CMB~\cite{Chang:2022tzj}, including those with Simons Observatory~\cite{SimonsObservatory:2018koc}, CMB-S4~\cite{Abazajian:2019eic,Snowmass2021:CMBS4}, PICO~\cite{NASAPICO:2019thw}, and CMB-HD~\cite{Sehgal:2019ewc,CMB-HD:2022bsz} will provide exquisite measurements of CMB temperature and polarization anisotropies across a wide range of scales.  These experiments will also provide maps of the integrated mass density through measurements of CMB lensing.  These measurements, especially when combined with galaxy surveys including those from DESI~\cite{DESI:2016fyo}, LSST~\cite{LSSTScience:2009jmu}, MegaMapper~\cite{Schlegel:2019eqc}, and PUMA~\cite{PUMA:2019jwd} will provide measurements of the energy content and large-scale structure growth at unprecedented precision, allowing for complementary constraints on physical effects that impact ultra-small scale structure.

The strength of this research program relies on its multiprobe nature and the synergies that the different facilities provide, from new GW detectors~\cite{snowmass_GWfacility}, to next generation telescopes, pulsar timing arrays~\cite{snowmass_facility} and CMB experiments~\cite{Chang:2022tzj}.

\section {\bf Dark matter physics probed by ultra-small-scale structure} 
In this section we provide more details about specific dark matter scenarios for which ultra-small-scale structure will provide key data: dissipative dark matter, primordial black holes, and ultralight bosons.

\subsection{Dissipative dark matter and dark black holes} 

While all observations are so far largely consistent with the cold dark matter (CDM) scenario of a single particle species weakly interacting with the standard model, there are observational hints from structure within galaxies for a more complex dark sector with interactions between dark matter particles. The lack of a definitive detection of dark matter in laboratory experiments constrains the possible ways dark matter may interact with the standard model. Together with the discovery of the Higgs but no other new particles at the Large Hadron Collider, these observations are spurring a shift in the theoretical landscape of possible physics beyond the standard model. New notions of naturalness, for example, suggest dark sectors that consist of relatively low mass particles decoupled or very weakly coupled to the standard model. This may be the start of a paradigm shift in dark matter studies away from a focus on an interaction between the dark matter and visible matter as cosmologically most important and toward a focus on possible interactions within the dark sector itself. 

The rich astrophysical structures observed electromagnetically form due to interactions that allow energy transfer between particles with very different masses and interactions. For example, the early stages of structure formation depend crucially on the dissipation of kinetic energy in a gas of hydrogen and electrons, for example via collisional excitation of the hydrogen atoms and molecules. The hydrogen then emits photons that can escape the gas entirely, allowing the gas to cool and condense enough for proto-stars to form. Dark matter with a particle spectrum more complex than that of a traditional CDM may similarly have dissipative interactions that become relevant only at some densities and temperatures so that the halo structure observed on large scales is maintained but structure can be clumpier on small scales \cite{Buckley:2017ttd}. A dramatically detectable example occurs if some dark matter gas can cool sufficiently for gravitational collapse to proceed. In the absence of nuclear physics, end state of collapse is a black hole. In a simple model, atomic dark matter \cite{Kaplan:2009de}, estimates have shown that if even 0.01\% of dark matter has collapsed into black holes, this population is accessible with current and near-future detectors \cite{Shandera:2012ke} and is likely to contain sub-solar mass objects. On the other end of the mass spectrum, similar scenarios may enable the formation of super-massive black holes \cite{DAmico:2017lqj, Latif2019,Chang:2018bgx}. Between these perhaps most detectable extremes, dissipative dark matter would likely impact the entire black hole mass spectrum. In addition, other possible novel compact objects, including neutron stars \cite{Hippert:2021fch} or white dwarfs \cite{Ryan:2022hku} made of dark matter. These objects would in general follow mass-radius or mass-compactness curves distinct from standard model objects and so would have distinct GW signatures.

\begin{figure*}[htb]%
	\centering
    \includegraphics[width=0.5\textwidth]{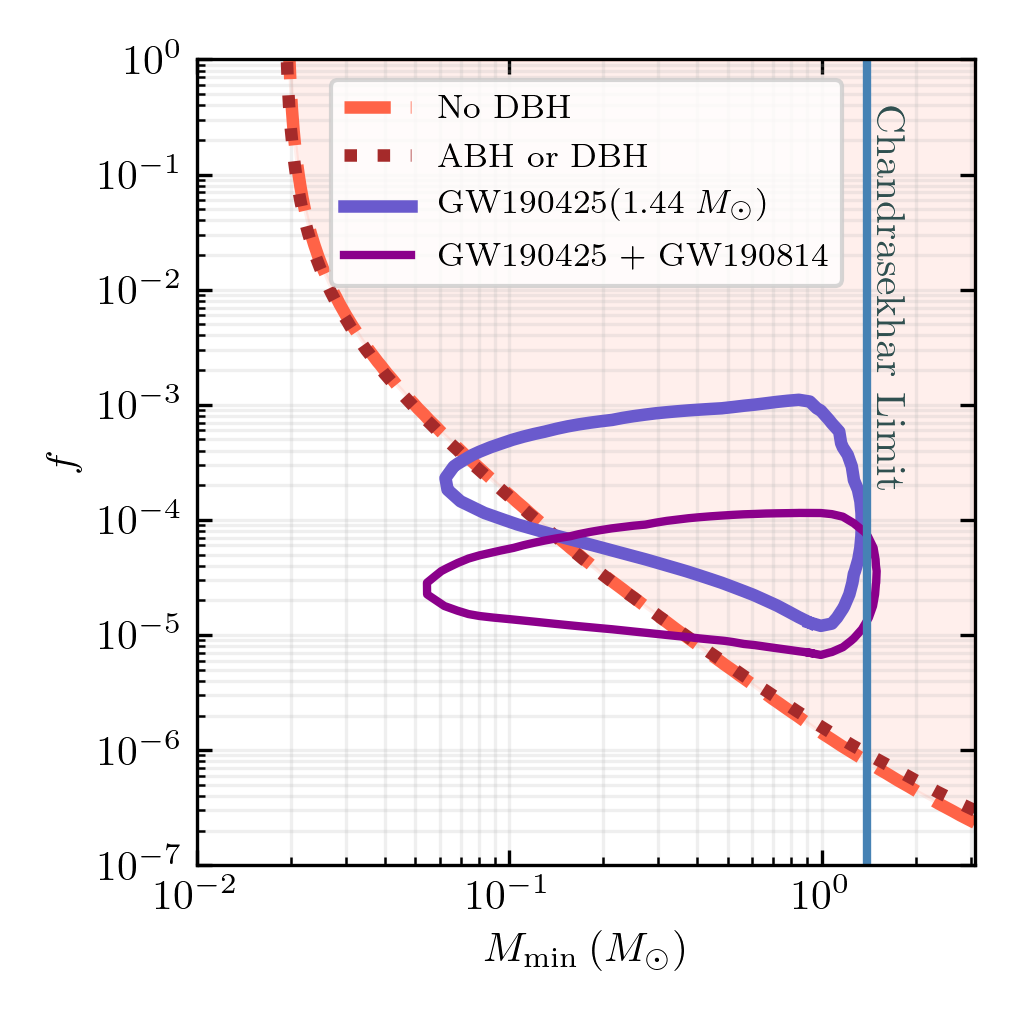}
	\caption{Constraint on the fraction of dark matter $f$ in dark black holes against the minimum allowed mass of the dark black hole $M_{\textrm{min}}$ for the dissipative dark matter model of~\cite{Shandera:2018xkn} if (i) none of LVC's binary black holes are dark black holes (\textit{dashed}), (ii) all LVC detections are either from astrophysical or dark black hole binaries with equal probability (agnostic about origin) (\textit{dotted}), (iii) GW190425 is a dark black hole binary detection (\textit{blue}), and (iv) GW190425 and GW190814 are from dark black hole binaries (\textit{pink}). Constraints for (iii) and (iv) are computed under an agnostic assumption about the origins of other LVC events. The Chandrasekhar limit for conventional black holes ($1.4~M_\odot$)
	is plotted for reference. The contours mark 90\% confidence regions. Figure from in Ref.~\cite{Singh:2020wiq}.
    \label{fig:DDMrates}
    }
\end{figure*}

On the sub-solar mass end, observations have the potential to illuminate dark matter microphysics in at least two straightforward ways. First, the smallest mass of any detected black holes provides a lower limit on the mass of fundamental particles in the dark sector via the Chandrasekhar bound. Second, the formation of  black holes from dark matter requires that the dark gas can cool to a low enough temperature for gravitational collapse to win over pressure. The detected mass of any dark black holes therefore bounds the cooling function, and the bound state energy levels (atomic or molecular) that enable cooling. These ideas were explored in detail in \cite{Singh:2020wiq}. Figure \ref{fig:DDMrates} shows the constraints on the fraction of dark matter in black holes as a function of the minimum mass of a potential dark black hole population, assuming none of the events detected Sept 12, 2015 - Oct 1, 2019 have a dark matter origin. Also shown are are the most likely regions of parameter space if GW190425 alone (purple) or GW190425 and GW190814 (pink) are mergers involving dark black holes. 

\begin{figure*}%
	\centering
\includegraphics[width=0.6\textwidth]{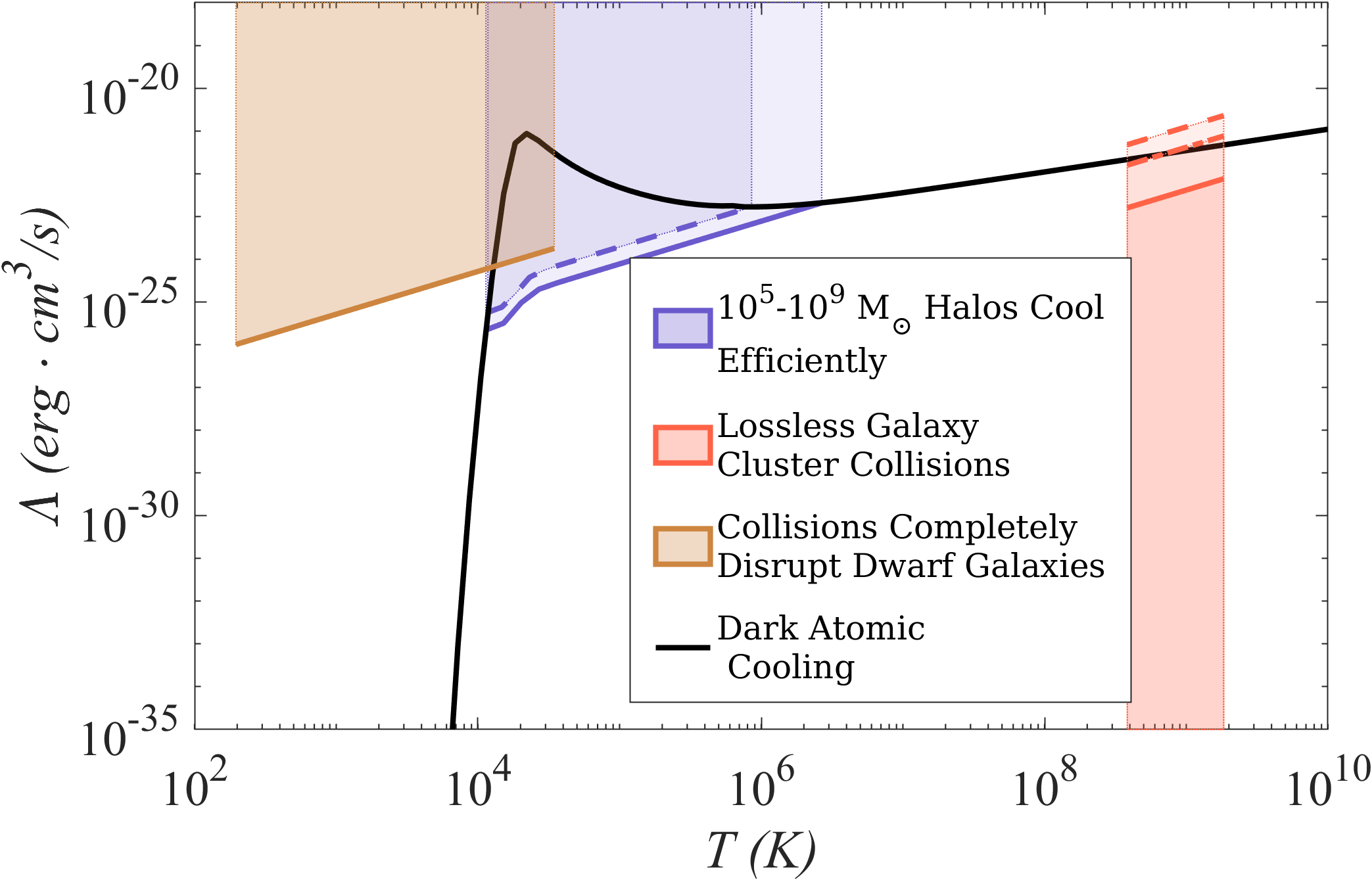}
	\caption{Cooling rate constraints from dark black hole detections combined with large scale structure collisions. Shaded regions correspond to the allowed parameter space in which (\textit{blue}) small halos at $z = 5, 10$ where dark black holes form, (\textit{red}) energy loss fraction $f_{\rm lost} = 0.01, 0.1, 0.3$ (solid to dashed), for galaxy cluster collisions, and (\textit{brown}) $f_{\rm lost} = 0.5$ for dwarf galaxy collisions. We also include the cooling function (solid black line) for an example atomic dark matter model consistent with the interpretation that GW190425 is a binary dark black hole $(m_x = \SI{14}{\giga\electronvolt}, m_c = \SI{325}{\kilo\electronvolt}, \alpha_D = 0.01)$. Figure from in Ref.~\cite{Singh:2020wiq}.
	\label{fig:coolingrate}
    }
\end{figure*}

Figure \ref{fig:coolingrate} shows how the large-scale constraints on interacting dark matter can be used together with ultra-small scale constraints from compact objects to constrain dissipation in the dark sector. The mapping between the dark matter microphysics and the black hole spectrum used in \cite{Singh:2020wiq} will likely be revised now that the molecular processes have been calculated \cite{Ryan:2021dis} and numerical tools are being developed \cite{Gurian:2021qhk, Ryan:2021tgw} to study a cooling gas of dissipative dark matter in detail. 

Dissipative dark matter provides a motivation to pursue GW data at high frequencies in order to constrain a potential population of sub-solar mass black holes. It is also a motivation to search existing data for binary mergers involving sub-solar mass compact objects. This analysis has been carried out by the LIGO/VIRGO/KAGRA collaborations and by independent groups ~\cite{Magee:2018opb,Abbott:2018oah,Authors:2019qbw,Nitz:2020bdb,LIGOScientific:2021job,Nitz:2021mzz,Nitz:2021vqh}. Although GW wave observations may not distinguish a dark black hole from a primordial black hole (discussed next), other aspects of small-scale structure will differ significantly between dissipative dark matter and primordial black hole dark matter. Clearly, a suite of small-scale observations are required to uncover the nature of dark matter. Finally, further work on a wider variety of dissipative dark matter scenarios, and the spectrum of black holes, neutron stars~\cite{Hippert:2021fch}, and white dwarfs~\cite{Ryan:2022hku} that can form from dark matter should inform analysis strategies for future GW observatories. See \cite{snowmass_DMextreme} for an additional discussion of GW signatures of dark sectors, and \cite{Bechtol:2022koa} for a broader discussion of how halo measurements can constrain dark matter.

\subsection{Primordial black holes}

Primordial black holes have long been considered as perfect Cold Dark Matter (CDM) candidates \cite{Carr:1975qj}, although their present renaissance has come in hand with the detection of GWs by the LIGO-Virgo interferometers from the merging of massive black holes that could be primordial \cite{Bird:2016dcv,Clesse:2016vqa,sasaki2016primordial}. We still do not know whether there are indeed primordial black holes among the LVK events \cite{Clesse:2020ghq}, and we may have to wait for a subsolar mass candidate, but even then we still could not tell if they comprise all of the dark matter \cite{Garcia-Bellido:2017fdg}. 

What has become clear in the last few years is that the formation of primordial black holes in the early universe is rather generic \cite{Garcia-Bellido:1996mdl}. We do not need unexpectedly large amplitudes of matter fluctuations but rather large non-Gaussian exponential tails, probably arising from the phenomenon of quantum diffusion during inflation, an effect overlooked until recently \cite{Ezquiaga:2019ftu}. Moreover, even the known thermal history of the universe may play an important role by providing the required lack of pressure to allow gravitational collapse at certain well defined epochs in the evolution of the universe \cite{Carr:2019kxo}. These are the Electroweak and QCD epochs at the time of $e^+ e^-$ annihilation, which generate a multimodal primordial black hole mass function with peaks at $10^{-5},\ 1,\ 10^{2},\ 10^{6}\ M_\odot$. These black holes come with different fractional abundances depending on the underlying inflationary potential and this may be used as a window into the early universe and fundamental physics. For example, if an excess of $10^{-10}\ M_\odot$ is found in microlensing events, or in the induced Stochastic Gravitational Wave Background (SGWB) at LISA frequencies, we may infer the existence of new fundamental particles at scales above those reached by present particle physics accelerators, which become non-relativistic and momentarily decrease the radiation pressure at a time when the mass within the horizon is precisely that mass scale.

The possibility that primordial black holes may be lurking in the dark universe as building blocks of the Cold Dark Matter fluid is extremely attractive \cite{Clesse:2017bsw}. In fact, the non-Gaussian exponential tails mentioned above may give rise to enhanced clustering, which leads, after recombination, to a population of primordial black hole clusters, with intermediate masses of order $10^{6}\ M_\odot$, that could be searched for with the microlensing of quasars around clusters or the perturbations they induce on stellar tidal streams around our galaxy and Andromeda \cite{Montanari:2020gcr}. These clusters could explain where most of the mass in the halo of galaxies is, thus evading the microlensing limits coming from stars in LMC and the bulge \cite{Calcino:2018mwh}, which had been used in the past to rule out primordial black holes as the main component of CDM \cite{Wyrzykowski:2019jyg}. Moreover, such dense objects may help explain many unexpected correlations in the radio and X-ray backgrounds at high redshift \cite{Kashlinsky:2016sdv,Kashlinsky:2019kac}, as well as the unusually high number of massive galaxies and quasars at high redshift, while the standard $\Lambda$CDM scenario could not account for them.

There is nowadays a great opportunity for testing all these ideas with new astrophysical and cosmological observations. For example, if primordial black holes existed before recombination, they should have left their imprint in an excess of injected energy in the plasma in the form of spectral distortions at high frequencies that a CMB experiment dedicated to it could detect \cite{Chluba:2019kpb}. One could further use the James Webb Space Telescope (JWST) to look for the first stars and galaxies at redshifts bigger than ten or twenty, confirming the role of black holes in early star formation \cite{Hasinger:2020ptw,Cappelluti:2021usg}.

Moreover, with the advent of the next generation of GW observatories, like the Cosmic Explorer \cite{Evans:2021gyd} and the Einstein Telescope \cite{Maggiore:2019uih} on the ground, and LISA in space \cite{Barausse:2020rsu}, we should be able to reach black hole fusions at high redshift (above 100), where no plausible stellar evolution could have generated such a population, and therefore convincingly prove their primordial nature. A further hint at their primordial origin which would link their formation with the cosmic history is the discovery of the induced SGWB from second order perturbations of large amplitude fluctuations entering at the same time of the primordial black hole formation \cite{Garcia-Bellido:2017aan}, when the size of the horizon redshifted today gives LISA frequencies (mHz), or perhaps in the PTA range of nHz. Such a discovery would open a new window into the early universe, where we could explore the non-Gaussian character of the fluctuations giving rise to the primordial black hole mass spectrum, as well as the number of relativistic species present at that time \cite{Carr:2019kxo}, well beyond the present reach with particle accelerators.

For reference, it is worth pointing out the Snowmass paper on Primordial Black Holes, Ref.~\cite{snowmass_pbh}, which specifically focuses on these natural candidates for dark matter, describing the science cases (the origin of primordial black hole dark matter in the early Universe) and the existing observational constraints, expanding on the theoretical work and data analysis required to improve the constraints and/or enable a possible detection in the future.

\subsection{Ultralight bosons and black hole superradiance}

Ultralight boson particles, including scalar (spin 0), vector (spin 1), and tensor (spin 2) fields, have been predicted under several theoretical frameworks to solve problems in particle physics, high-energy theory and cosmology \cite{Peccei1977,Peccei1977PhRvD,Weinberg1978,Arvanitaki2010,Goodsell2009,Jaeckel:2010ni,Essig:2013lka,Hui:2016ltb}. 
One well-motivated example is the quantum-chromodynamics (QCD) axion, a pseudo-Goldstone scalar boson proposed to explain the strong charge-parity (CP) problem~\cite{Peccei1977,Peccei1977PhRvD,Weinberg1978}.
Other examples include a variety of axion-like particles with masses in a range spanning between $10^{-33}$ eV and $10^{-10}$ eV (e.g., string axiverse~\cite{Arvanitaki2010}) and ultralight vector particles (e.g., hidden photons arising from string theory~\cite{Goodsell2009}). 
These bosons could also be a significant component of dark matter~\cite{Jaeckel:2010ni,Essig:2013lka,Hui:2016ltb}.
	
These proposed new particles are difficult to detect by conventional particle physics experiments, due to their extremely small mass and weak interaction (if any) with Standard Model particles.
For example, existing constraints on the existence of the QCD axion and its mass ($\lesssim 10^{-3}$ eV) obtained from conventional experiments are based on particular assumptions of its expected coupling to the Standard Model~\cite{Kim:2008hd,Raffelt2006}.
However, as predicted by theory, a mass range of $\lesssim 10^{-10}$~eV is favored for the QCD axion~\cite{Arvanitaki2011}.
For such conjectured bosons with ultralight mass and vanishing interaction with the Standard Model, gravitational coupling would be the only possible way to study them.

Around a rapidly spinning black hole, if there exists a fundamental ultralight boson field, it is expected to grow when the superradiance condition is satisfied~\cite{Penrose1969,Press1972,Bekenstein1973,Arvanitaki2010,Brito2014,Brito2015}: $\omega_\mu / m < \Omega_{\rm BH}$, where $\omega_\mu=\mu/\hbar$ is the characteristic angular frequency of a boson with rest energy $\mu$, $m$ is the boson azimuthal quantum number with respect to the rotation axis of the black hole, and $\Omega_{\rm BH}$ is the angular speed of the black hole's outer horizon.
For a black hole with mass $M$, when the Compton wavelength of the particle is comparable to the characteristic length of the black hole, i.e., \mbox{$\hbar c/\mu \sim GM/c^2$}, the superradiant instability is maximized, and thus the occupation number of ultralight bosons grows exponentially, forming a macroscopic cloud comprising ${\sim}10\%$ of the black hole mass. After forming, the bosons in the cloud can annihilate, which would generate continuous, quasimonochromatic GWs over a long lifetime, which can be observed by GW detectors~\cite{Arvanitaki2011,Yoshino2014,Yoshino2015,Arvanitaki2015,Arvanitaki2017,Brito2017-letter,Brito2017,Baryakhtar2017,Siemonsen2020,Brito2020}. 

In this paper, we focus on the GW searches for ultralight boson clouds. More discussions are detailed in Ref.~\cite{snowmass_DMextreme} about other studies of superradiance, including the inference of the boson existence from the spin distribution of black holes with next-generation GW detectors, numerical and theoretical opportunities in superradiance, etc.


\subsubsection{Astrophysical sources}

Primary types of individual sources for ground-based detectors include: (1) remnant black holes from compact binary coalescences (CBCs) \cite{LVC-catalog}, (2) known black holes in X-ray binaries \cite{Remillard:2006fc,Yoshino2015,Middleton2016}, and (3) isolated black holes in the Milky Way.  
	
Nearby remnant black holes from detected CBC events are ideal targets~\cite{Arvanitaki2017,Baryakhtar2017,LVC-catalog}.
The age of the newly born remnant black hole is perfectly known, and the black hole's intrinsic parameters (mass and spin) can be inferred through the CBC parameter estimation. This enables accurate theoretical predictions of the GW signal waveform from the boson cloud around the remnant black hole. 
Moreover, the extrinsic parameters, e.g., the sky position and orientation of the source, can also be well measured for a sufficiently loud CBC event, allowing us to conduct a dedicated, efficient follow-up search for the conjectured cloud in the detector data. Detection prospects for CBC remnants using current-generation detectors are penalized by the typically large luminosity distances. However, future detectors promise to enable stringent boson constraints and/or possibly a detection over significantly increased luminosity distance. 
	
Known black holes in X-ray binaries~\cite{Remillard:2006fc,Middleton2016} are much closer and hence potentially within the detectable range of existing detectors.
Constraints on the boson mass have already been suggested from spin measurements of black holes in X-ray binaries, roughly disfavoring the mass range of $10^{-12} \lesssim \mu/{\rm eV} \lesssim 10^{-11}$ for axion-like scalars \cite{Arvanitaki2015,Cardoso2018,Stott:2018opm} and $10^{-13} \lesssim \mu/{\rm eV} \lesssim 10^{-11}$ for vectors \cite{Baryakhtar2017,Cardoso2018}. Direct probe though GW searches would allow for dedicated in-depth study of these systems. 
There are, however, some challenges associated with these sources that need to be taken into account: 
The age and history of these systems are largely uncertain; 
there might be impact from the systematics in the black hole spin measurements~\cite{Reynolds:2013qqa,McClintock:2013vwa}; 
the active astrophysical environments surrounding these black holes needs to be better understood~\cite{Arvanitaki2015,Baryakhtar2017,Baumann:2018vus}.
In addition, the Doppler modulation due to the black hole motion within the binary needs to be accounted for, increasing the complexity of these searches.
	
Isolated black holes in the Milky Way are also extremely interesting~\cite{Arvanitaki2015,Arvanitaki2017,Brito2017-letter,Brito2017}, due to their expected abundance and proximity to Earth, and as they could provide a cleaner signal compared to the X-ray black holes. The lack of electromagnetic counterparts, however, requires searching over a larger parameter space (with reduced sensitivity) since in general no prior information of the black hole position and parameters is available.  

In addition to the above sources, with the launch of space-based GW detectors (e.g., LISA \cite{LISA}, TianQin \cite{TianQin}) in the next decade, the boson mass range of \mbox{$10^{-19}\lesssim \mu/{\rm eV} \lesssim 10^{-15}$} will also become accessible to GW detectors through the same superradiance phenomenon, corresponding to boson clouds grown around massive black holes with masses in the range $\sim 10^3$--$10^8 M_\odot$~\cite{Brito2017}. 

\subsubsection{Gravitational-wave searches}

Searches have already been carried out with existing ground-based GW detectors. 
The search for GW signals from boson clouds around spinning black holes is conceptually similar to the ``standard" searches for continuous waves (CWs) from individual, asymmetric spinning neutron stars~\cite{Riles2017,Prix2009,Sieniawska2019}.
We can take advantage of the existing methods and framework developed for standard CW searches, and apply them to searching for signals from individual boson clouds. 

A semi-coherent, computationally efficient method using a collection of fast Fourier transforms (FFTs) computed over various durations (from hundreds to thousands of seconds), has been designed and implemented for all-sky CW searches~\cite{DAntonio2018}. 
An estimate of the detection reach for vector boson condensates, derived from all-sky CW searches in the first Advanced LIGO observing run (O1) has been presented in~\cite{Dergachev2019}.
The first constraints on the scalar boson mass (in a range $\sim 10^{-13}$\,eV) have been derived in~\cite{Palomba2019}, using strain upper limits obtained from all-sky CW searches in the second observing run of Advanced LIGO (O2).
The black hole population is considered for characterizing ensemble signals from galactic isolated black holes~\cite{Zhu:2020tht}.
Ref.~\cite{Isi2019} modeled the signal waveforms for individual sources, and demonstrated the suitability of a specific search algorithm based on a hidden Markov model to efficiently search for signals from a known sky position~\cite{Suvorova2016,ScoX1ViterbiO1,Sun2018}.
A dedicated search for GWs from ultralight scalars in Cygnus X-1 was conducted using Advanced LIGO O2 data~\cite{Sun2020}, with and without considering the nonlinear self-interaction of the bosons~\cite{Arvanitaki2010,Arvanitaki2015,Yoshino2012,Yoshino:2015nsa,Yoshino2015}. Direct constraints on the boson mass and decay constant have been obtained~\cite{Sun2020}.
In the third observing run of Advanced LIGO and Virgo (O3), an all-sky search for tailored for scalar bosons has been carried out, yielding some of the most stringent upper limits on the existence of boson clouds around spinning black holes in the Milky Way~\cite{O3boson}. 

Finally, in addition to individual sources, searches for a stochastic GW background from scalar and vector boson cloud signals have also been proposed and carried out~\cite{Brito2017-letter,Tsukada2019,Tsukada2021}.

\subsubsection{Future prospects}

\begin{figure*}%
	\centering
	\subfloat[Cosmic Explorer\label{fig:horizon_ce}]{\includegraphics[width=0.5\textwidth]{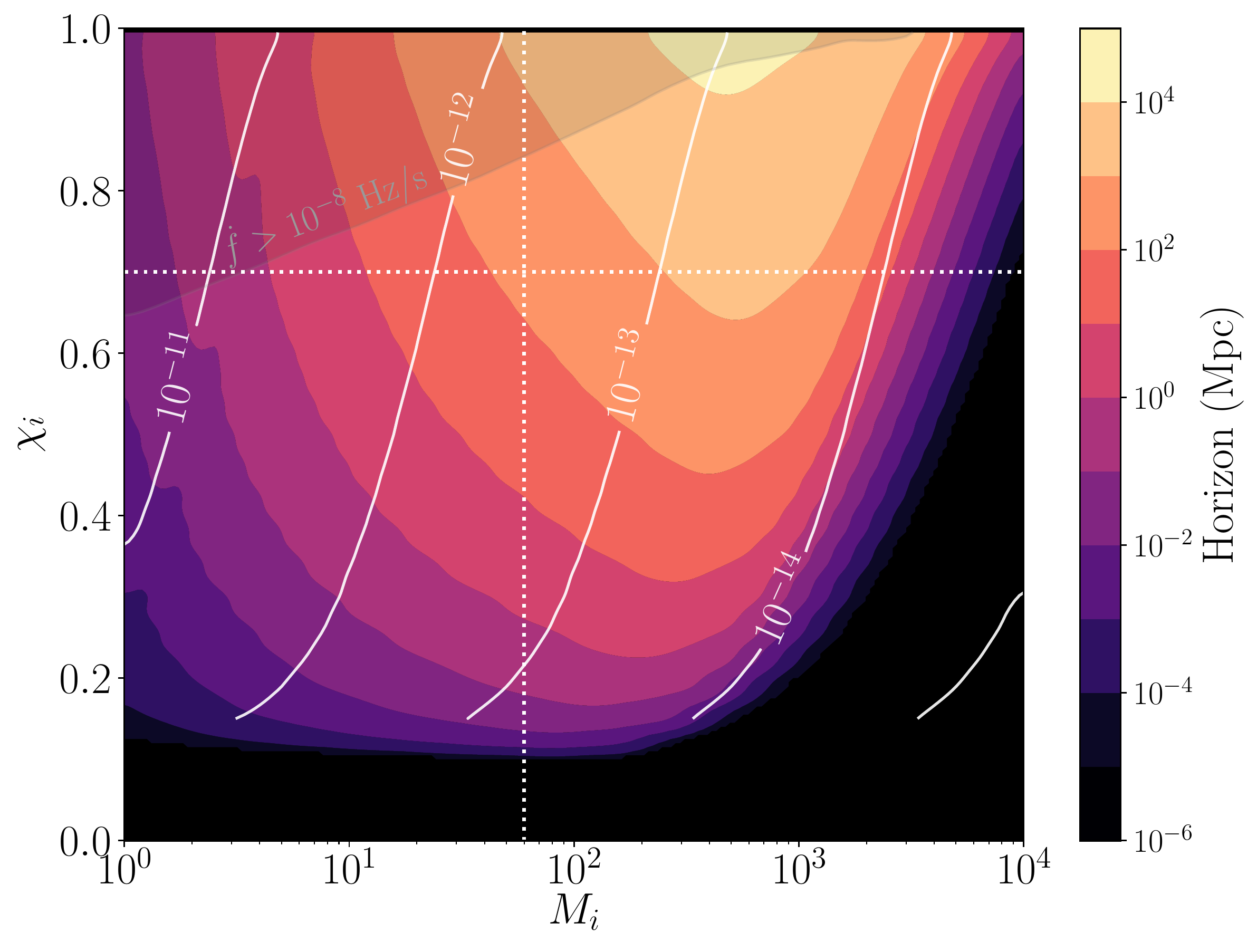}}
	\subfloat[Einstein Telescope\label{fig:horizon_et}]{\includegraphics[width=0.5\textwidth]{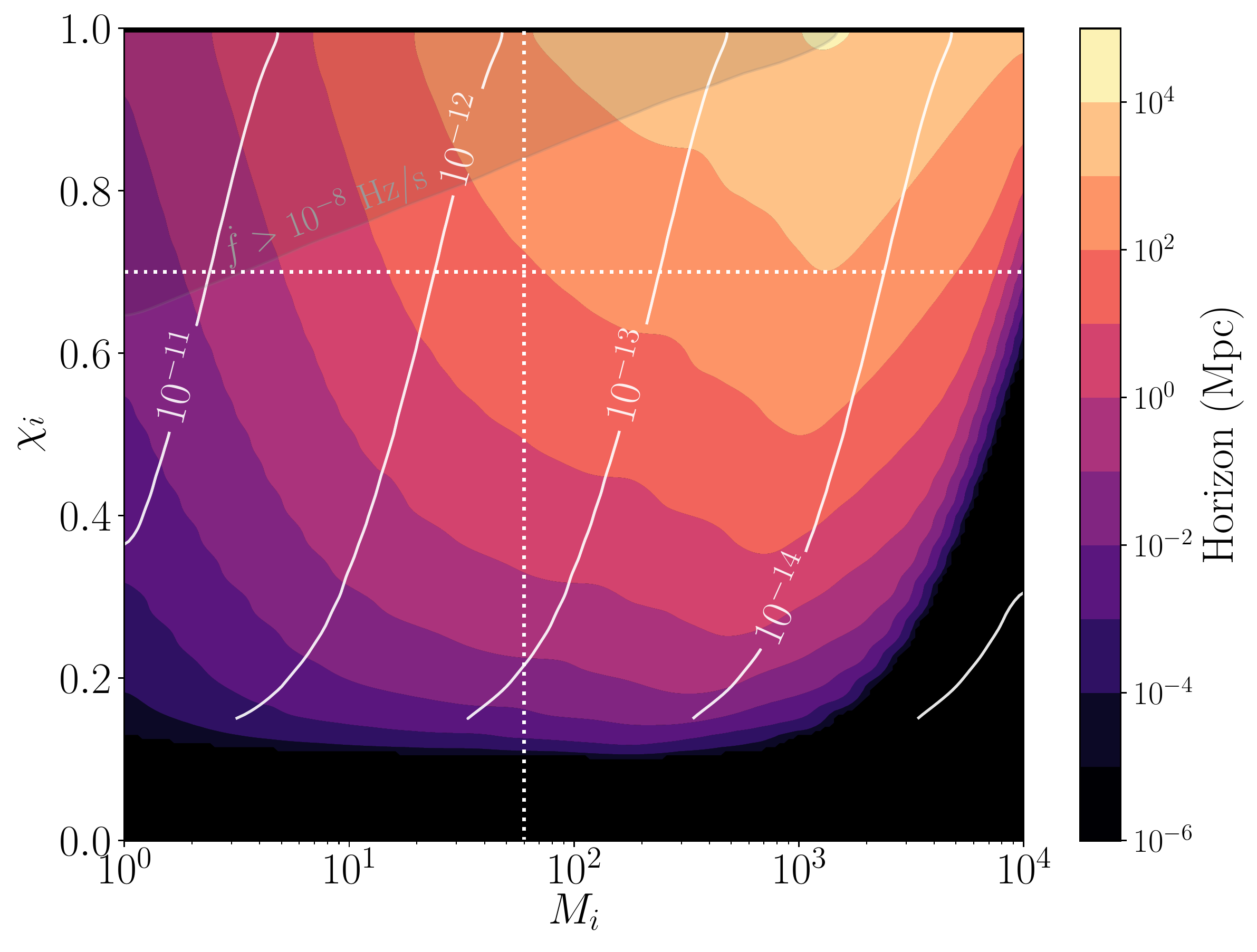}}
	\caption{(Fig.~12 in Ref.~\cite{Isi2019})
    Maximum detectable luminosity distances (color) for optimal scalar clouds around black holes with initial mass $M_i$ and spin $\chi_i$ for (a) Cosmic Explorer and (b) Einstein Telescope (one year of observation by a single detector). White contour lines indicate the values of the corresponding boson rest-energy $\mu/{\rm eV}$. }
	\label{fig:horizons}
\end{figure*}

The next-generation ground-based detector network, including Einstein Telescope \cite{Punturo:2010zz,Hild:2010id} and Cosmic Explorer \cite{nextGen-2017,Reitze2019cosmic,Evans2021horizon}, as well as space-based GW detectors will promise the capability of reaching a significantly increased luminosity distance and probing a large parameter space of the ultralight bosons. In particular, follow-up searches for CBC remnant black holes in the latest observing runs that are still limited by the detectable luminosity distance with existing detectors, will reach a detectable luminosity distance up to $\sim 10^4$~Mpc for scalars with mass $10^{-14}\lesssim\mu/{\rm eV}\lesssim 10^{-13}$ with the next-generation detectors (e.g., see Fig.~\ref{fig:horizons}) \cite{Isi2019,Ghosh:2018gaw,Hild:2010id,Sathyaprakash:2012jk,Punturo:2010zz,Abbott2017-nextGen-CE}.
In addition to remnant black holes from CBC events, individual pre-merger black holes in compact binary systems are also proposed to be interesting CW sources for ground-based detectors, when those inspirals become accessible to the LISA detector, allowing for the measurement of the source properties~\cite{Ng2020}. Finally, given that LISA will be sensitive to GWs with much lower frequencies than ground-based detectors, it will be able to probe the complementary mass range \mbox{$10^{-19}\lesssim \mu/{\rm eV} \lesssim 10^{-15}$}~\cite{Brito2017,Brito2017-letter}.

On the other hand, progress has been made in theoretical studies of signals from vector and tensor fields.
The signal waveforms have been computed numerically from black hole perturbation theory~\cite{Siemonsen2020}, which largely facilitates the implementation of improved methods to look for vector bosons. Moreover, recent work presents the analytically computed GW emission from superradiant instability triggered by massive tensor fields and predicts the parameter space that can be probed through GW observations~\cite{Brito2020}, setting up the theoretical basis for empirical studies of tensor fields around black holes using GW data.
Signals from vector and tensor fields are expected to be stronger due to the intrinsically higher radiated power, which in principle could increase the distance to the detectable sources~\cite{Baryakhtar2017,Brito2020,Brito2020handbook}. However, the detectability is penalized by a much shorter signal lifetime, also due to the higher radiated power. Nonetheless, theoretical advances continue to enable new studies of vector and tensor fields, and the detection chances are expected to be improved with next-generation detectors and with LISA. 

\section{Gravitational-wave interferometers as dark matter detectors}

GW detectors could directly detect DM that interacts with various interferometer components. While the signal embedded in the detector noise would not arise from GWs, a detection would still offer new and interesting insights into the nature of ultralight DM, just as conventional direct DM detection experiments do.


\subsection{Sources}

Different types of DM would cause fundamentally different differential strains in current and future GW detectors. Scalar, dilaton DM would cause time-dependent oscillations of the values of fundamental constants, such as the electron mass \cite{Stadnik2015a,Stadnik2015b,Stadnik2016}. Physically, the Bohr radius would change, causing time-varying changes in the size and index of refraction of the beam splitter \cite{grote2019novel}. Since the light from each cavity would traverse a slightly different path on the surface of the beam splitter, a differential strain would result whose magnitude is independent of the length of the arms, and what matters is the amount of quantum noise reduction; thus, the most sensitive detector is actually GEO600, since it employs squeezed light vacuum light that greatly reduce quantum noise compared to LIGO/Virgo/KAGRA. Additionally, axions could couple to the laser light and alter the phase velocities of left- and right-hand circularly polarized light \cite{nagano2019axion}. In this case, the birefringence in the interferometer, i.e. optical path difference
between p- and s-polarized lights, has to be measured using some additional optics \cite{nagano2021axion}. 

Vector dark matter, such as dark photons arising from e.g. the misalignment mechanism \cite{nelson2011dark} or cosmic string network decays \cite{long2019dark}, would interact with baryons in the input and end mirrors, causing oscillatory forces on them that can be formulated as arising from a ``dark electric field'', analogously to the ordinary photon \cite{Pierce:2018xmy}. Since the dark matter field sees each of the mirrors in a different location with respect to its propagation direction, each one experiences a slightly different force, leading to different travel times for light down each arm and hence a differential strain \cite{Pierce:2018xmy}. Furthermore, an additional contribution to the differential strain arises due to the finite amount of time light takes to traverse each arm, a ``common-mode motion'' effect \cite{morisaki2021improved}. 
Tensor dark matter \cite{Marzola:2017lbt}, arising as a modification to gravity that could also play the role of DM \cite{Aoki:2017cnz}, could also cause a differential strain analogously to GWs by stretching and squeezing the spacetime around the mirrors \cite{Armaleo:2020efr}.

Primordial black hole dark matter also leave signatures in GW detectors through their coalescence or scattering. The signal from primordial black hole binaries is rather distinct. It could arise from distant sources $(z>10)$, correspond to black holes in the (upper) Pair Instability Super Novae mass gap, $60 < M < 120\,M_\odot$, or the (lower) Neutron Star - Black Hole mass gap, $2 < M < 5\,M_\odot$, but most importantly a black hole with less than one solar mass (Chandrasekhar limit) would signal its non-astrophysical origin. Moreover, primordial black holes are typically born without spin, while stellar black holes are known to have large spins. A prediction of clustered primordial black holes is that the spin distribution of black hole binaries should be peaked around negligible projected spin and misaligned with the orbital angular momentum \cite{Garcia-Bellido:2020pwq}.

\subsection{Signals}

Though the physical motivations behind DM interactions with GW detectors differ, all types of ultralight DM share some common traits. First, the ultralight mass, coupled with the known local DM energy density, imply a gargantuan number of DM particles in a given volume of space, whose wavefunctions overlap and can therefore be modelled as a superposition of plane waves \cite{carney2019ultralight}. Since the DM is cold, the velocities of each individual DM particle likely follow a Maxwell-Boltzmann distribution, centered around the velocity at which DM orbits the center of the Milky Way, the so-called ``virial velocity'', $\sim 10^{-3}c$ \cite{smith2007rave}. The DM velocities and ultralight masses imply finite coherence lengths and coherence times of the signal, the former ensuring correlated signals in earth-based detectors since the coherence length greatly exceeds the detector separation; the latter implying perfectly sinusoidal oscillations of the field fixed by the mass if observing for less than a coherence time, but deviations from monochrome since the observation time is much longer than the coherence time \cite{Miller:2020vsl}. The detectors always exist within the DM field, ensuring interactions that last much longer than the signal coherence time and the need for techniques to handle stochastic changes in the signal frequency over the observation time.

If primordial black holes have a multimodal mass function, as in the thermal history scenario \cite{Carr:2019kxo}, and a significant fraction of them are associated with their production at the QCD scale, corresponding to primordial black hole masses of around a solar mass, then we should also see an important SGWB in present laser interferometers, but very likely in future ground- and space-based GW antennas \cite{Braglia:2021wwa}.

\subsection{Searches}

Multiple searches for DM directly interacting with GW detectors have been performed over the last few years. The methods employed rely semi-coherent approaches, in which the data are broken into small time segments and analyzed with the phase information, and then combined afterwards without the phase information. 

\begin{figure*}[tbh!]
	\centering
	\subfloat[\label{fig:scalar}]{\includegraphics[width=0.6\textwidth]{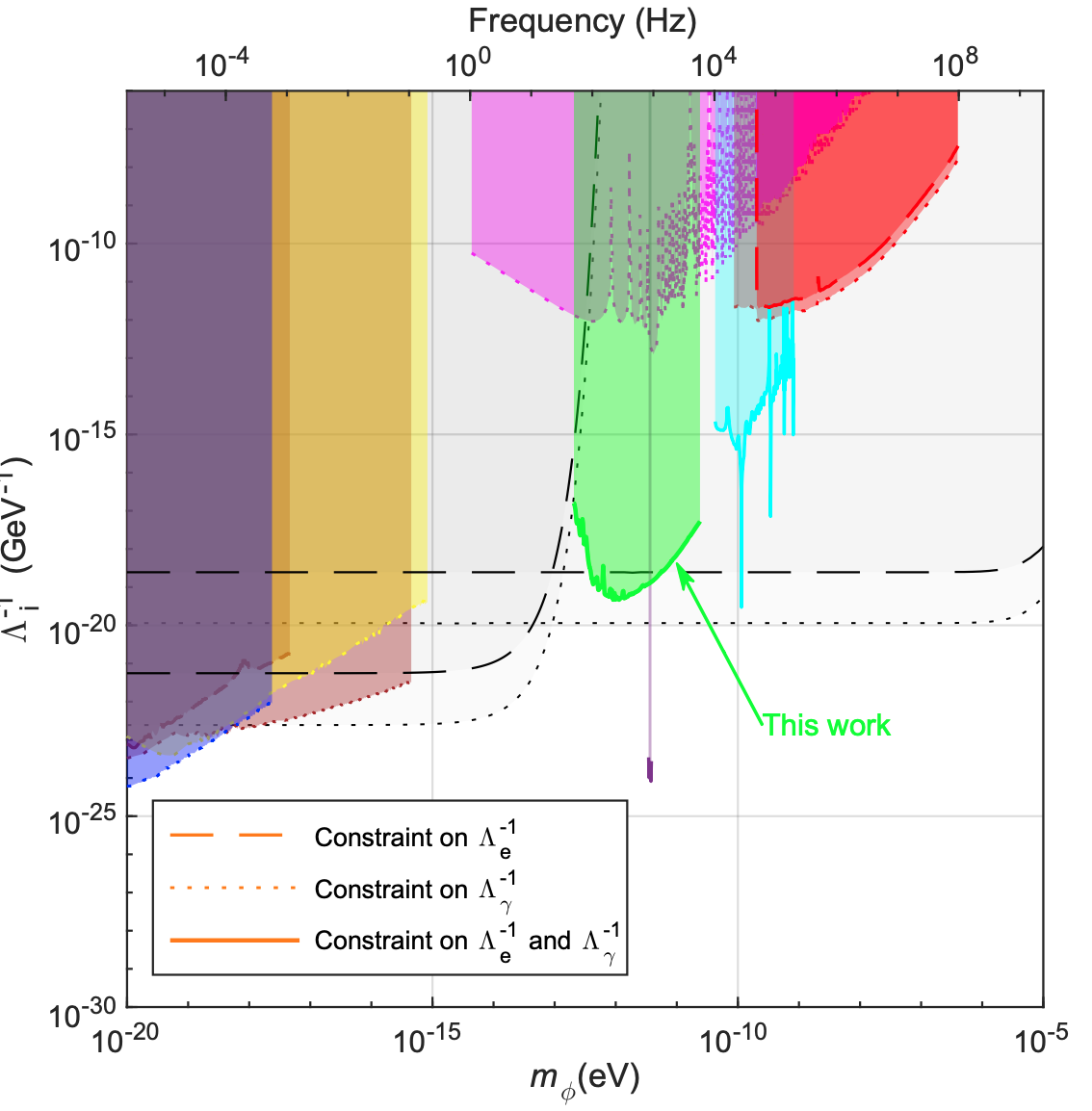}}
	\caption{(Fig. 2 in \cite{Vermeulen:2021epa}).
    Constraints on scalar, dilaton dark matter coupling to GEO600, which alters the electron mass and change the size and index of refraction of the beam splitter. }
\end{figure*}

In the search for scalar, dilaton DM with GEO600, an LPSD method was employed that varies the FFT length as a function of logarithmicly-spaced frequencies, ensuring optimal sensitivity to each DM mass \cite{Vermeulen:2021epa}. Such an optimization comes at the price of an increased computational cost, though this cost lies much below those of standard GW searches. Competitive upper limits were produced on three models for scalar DM that surpassed those from existing experiments, which can be seen in figure \ref{fig:scalar}.

\begin{figure*}%
	\centering
	\subfloat[\label{fig:dpdm}]{\includegraphics[width=0.8\textwidth]{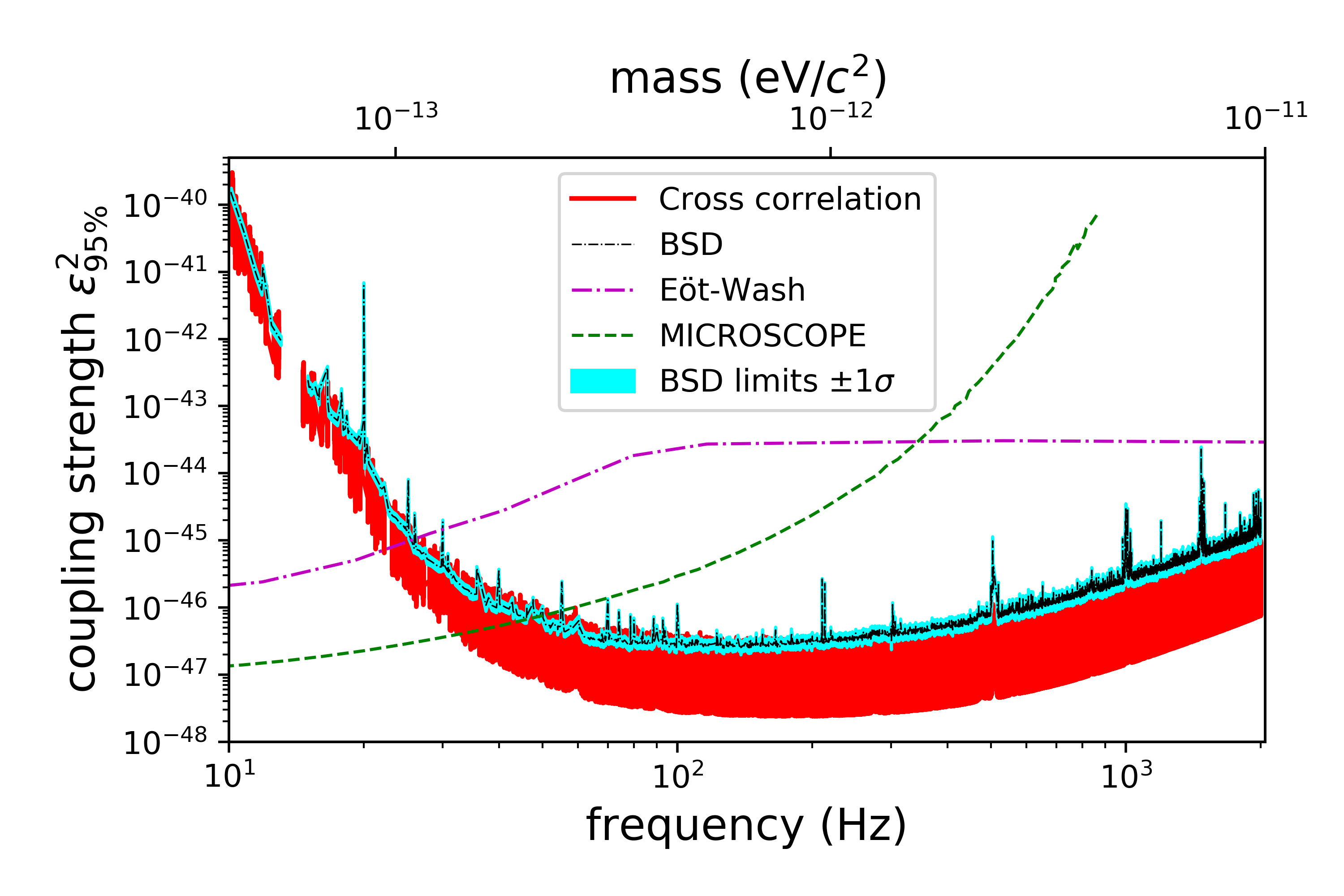}}
	\caption{(Fig. 3 in \cite{LIGOScientific:2021odm}).
    Constraints on the coupling strength of dark photons to baryons in the LIGO mirrors using two methods: cross-correlation and excess power (BSD). }
\end{figure*}

For dark photon DM, one technique relies on cross-correlation \cite{Pierce:2018xmy}, in which at least two data streams are analyzed together to look for a signal whose signal-to-noise ratio is negative. This search is very efficient, and produced the first-ever upper limits on dark photon DM coupling to baryons using data from the first observing run of LIGO \cite{guo2019searching}. Furthermore, a second method \cite{Miller:2020vsl} looks for excess power in individual detectors, then further analyzes coincident frequency bins that contain a high-enough SNR. Here, the FFT length is judiciously varied such that the frequency-dependent modulation due to the individual velocities of dark photons is confined to one frequency bin, which allows optimal sensitivity to each possible dark photon mass. While not as sensitive as cross-correlation, both methods have produced extremely stringent upper limits on dark photons using data from the third observing run of LIGO/Virgo \cite{LIGOScientific:2021odm}. These results improved upon existing constraints on the squared coupling of dark photons to baryons by a few orders of magnitude, and can be seen in figure \ref{fig:dpdm}. Finally, constraints on extremely ultralight DM have been placed with pulsar timing arrays, in which a cross-correlation analysis had to account for both correlated and uncorrelated signals depending upon the coherence length of the dark photon \cite{PPTA:2021uzb}. 

\subsection{Prospects}

The future of this field is bright. As current detectors improve their sensitivities and take data for longer periods of time, the limits on the coupling of DM to standard model particles will only become more stringent, allowing for a possible detection of ultralight DM. Most DM models do not predict lower bounds for the coupling strength; therefore, it is necessary to continue to probe the existence of DM in a wide mass range. Furthermore, projections for searches for tensor bosons and axions have been forecast \cite{Armaleo:2020efr,nagano2021axion}, which allow for the possibility of deep constraints with current and future ground-based GW detectors. In the case of axions, minimal hardware adjustments to the detection ports would allow a birefringence measurement and would not affect GW detection \cite{michimura2020ultralight}. Furthermore, sapphire mirrors could greatly improve the sensitivity of all GW detectors to vector dark matter that couples to baryon-lepton number \cite{michimura2020ultralight}, since this would enhance the number of charges in the mirrors relative to fused silica.

Looking further in to the future, space-based detectors, such as LISA, DECIGO and TianQin, will allow a different ultralight mass regime to be probed, $\sim 10^{-17}-10^{-15}$ eV/$c^2$, which will require the adaption of techniques to handle space detector data, as well as to work in a regime with potentially uncorrelated signals.

Also see Ref.~\cite{snowmass_stochastic}, which is specifically focused on probing early-universe GW and fundamental physics. Here, different forms of particle and macroscopic DM are discussed in the context of what their detections would mean for early-universe physics.

\section{Complementary and integrated probes of total matter content of the universe}


\subsection{Light relics}

Models that predict ultra-small-scale structure that differs from what is expected with standard cold dark matter often contain new light degrees of freedom.  One example is provided by axions, which may have a cold population acting as the dark matter as well as a hot component that contributes to the radiation density~\cite{Baumann:2016wac}.  Another example is provided by dark matter that undergoes self-interactions mediated by some new light species in the dark sector.  The density of light relics, as measured with primordial light element abundances, the CMB power spectrum, and large-scale structure, receives contributions from any particles that have the same gravitational influence as relativistic neutrinos, which is true of any (free-streaming) radiation~\cite{Green:2019glg}. See also the dedicated Snowmass 2021 White Paper~\cite{Snowmass2021:LightRelics} for a further discussion of the physics of light relics.

If these additional light species have any couplings to the Standard Model, they may have been in thermal equilibrium with the plasma filling the early universe.  At early times, the density and temperature of the plasma were extremely high, much higher even than in the cores of stars or supernovae.  Measurements of the light relic density thereby allow deep insights into the existence of new light species, even those that may have evaded detection in other searches.
It is straightforward to predict the thermal relic density of any light species that decoupled from the plasma while relativistic. Under mild assumptions~\cite{Wallisch:2018rzj}, the contribution to the effective number of neutrino species~$\Delta\Neff$ of any light thermal relic is determined by two numbers, the temperature at which the relic decoupled from the Standard Model plasma,~$T_F$, and its effective number of spin degrees of freedom,~$g_s$, specifically $\Delta\Neff = \frac{4}{7}g_s \left(\frac{43/4}{g_{\star}(T_F)}\right)^{\!4/3}$.
The function~$g_{\star}(T_F)$ is the effective number of relativistic degrees of freedom present in the thermal plasma at the temperature~$T_F$ (defined as the effective number of independent states including a factor of~$7/8$ for fermions).  The quantity $g_{\star}(T_F)$  encodes the dilution of the thermal relic relative to photons due to the annihilation of heavy states after the relic decouples from the plasma. The next generation of CMB~observations are expected to reach a precision of $\sigma(\Neff) = 0.03$~\cite{CMB-S4:2016ple, NASAPICO:2019thw, Abazajian:2019eic}, which would extend our reach in~$T_F$ by several orders of magnitude, thereby allowing detection of or severe constraints on new light species with very weak couplings to the Standard Model.

Measurements of the light relic density provide useful insights into thermal history of the universe, even in models that do not include new light species. Since GWs contribute to the light relic density, models that result in a stochastic background of GWs~\cite{Boyle:2007zx, Stewart:2007fu, Meerburg:2015zua} can be usefully constrained by measurements of $\Neff$, which acts as an integral constraint on the stochastic GW background. Violent phase transitions and other nonlinear dynamics in the primordial universe can lead to a large amplitude of GWs, with a spectrum that may be peaked at frequencies much larger than those accessible to $B$-mode polarization measurements of the CMB and in some cases at frequencies above the range probed by LIGO and~LISA~\cite{Maggiore:1999vm, Easther:2006gt, Dufaux:2007pt, Amin:2014eta,Caprini:2018mtu}. For particularly violent sources, the energy density in GWs can be large enough to make a measurable contribution to~$\Neff$~\cite{Caprini:2018mtu, Adshead:2018doq, Amin:2018kkg}.

Light relic particles do not need to be massless. Light relics with non-zero mass can have observable signatures in the large-scale structure of the universe.
A light (but massive) relic (LiMR) \cite{DePorzio:2020wcz} that is non-relativistic today will contribute to the dark matter density of the universe, but it will differ from dark matter in its temperature \cite{Munoz:2018ajr}. An example of a LiMR in the Standard Model is provided by neutrinos. Other examples include axions, dark photons, and gravitinos.

The decoupling of these relics while relativistic gives these particles significant streaming motion, which sets a scale below which they cannot cluster, altering the large-scale structure of the universe. In particular, they produce a suppression in the matter power spectrum at scales smaller than their free-streaming scale, with the size of this suppression depending on the present abundance of the LiMR, allowing the relic’s temperature and mass to be measured independently (see \cite{Xu:2021rwg} for details on a search on existing data).

\subsection{Neutrino mass constraints}

Over the past decade, we made significant progress in our understanding of the clustering of the matter field in the presence of massive neutrinos, reaching constraints tighter than any laboratory experiment. We expect this limits to improve with forthcoming experiments, reaching a detection of the absolute neutrino mass scale \cite{Dvorkin:2019jgs}.

\begin{figure}[thb!]
    \centering
    \includegraphics[width=5in]{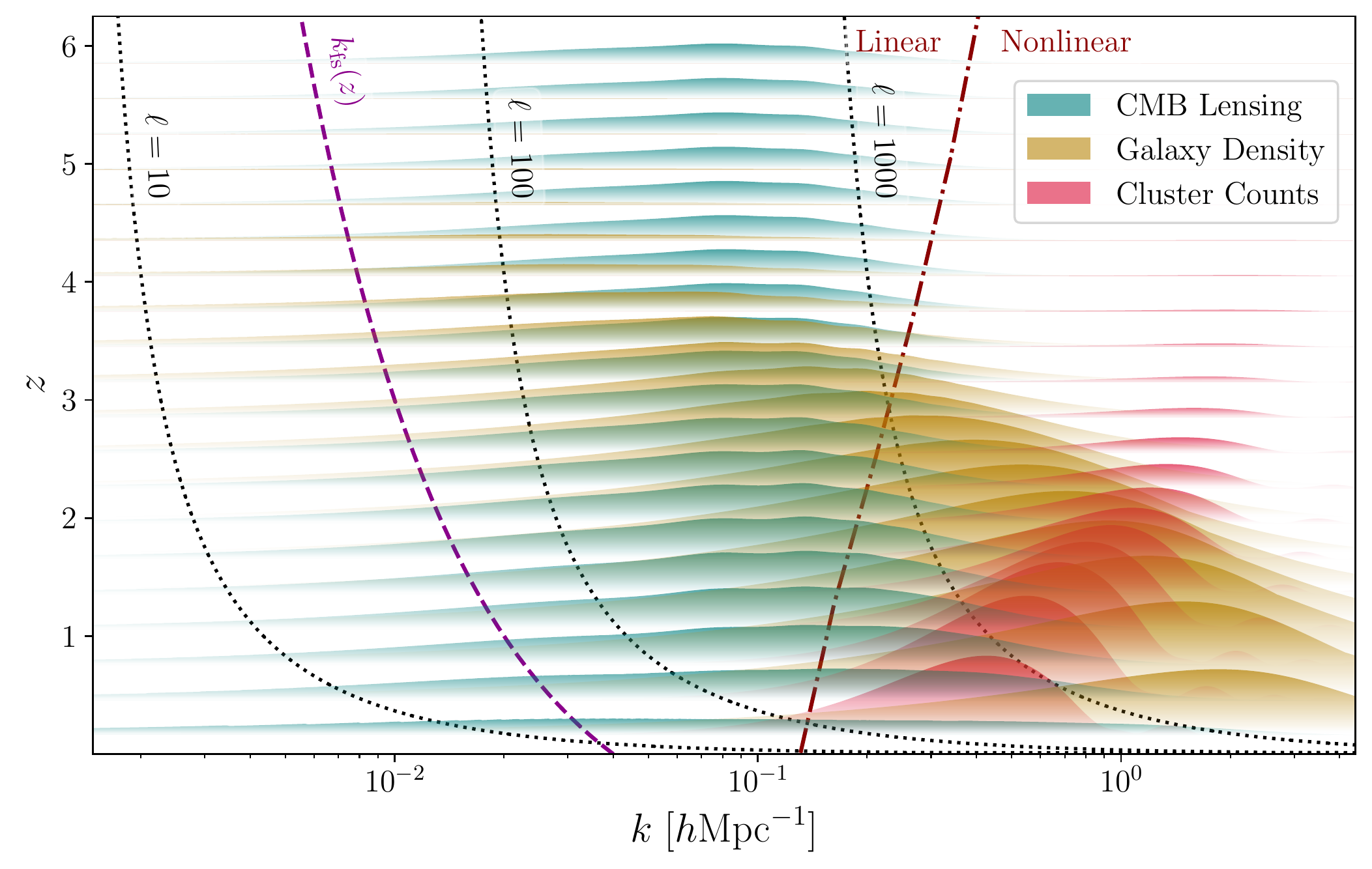}
    \caption{Contributions of the matter power spectrum $P(k,z)$ to various observational probes of clustering.  Heights denote the signal-to-noise ratio anticipated for each observable: the CMB lensing power spectrum using the lensing reconstruction expected from CMB-S4~\cite{CMB-S4:2016ple,Abazajian:2019eic,Snowmass2021:CMBS4}, the angular power of galaxy density using observations from the LSST gold sample~\cite{LSSTScience:2009jmu}, and number counts of clusters with mass greater than $10^{14}~h^{-1}M_\odot$ corresponding roughly to the detection threshold for the thermal Sunyaev-Zel`dovich effect anticipated with CMB-S4~\cite{CMB-S4:2016ple}.  The CMB lensing weighting is scaled up by a factor of 3 relative to the others in order to make the CMB lensing contributions more visible in the displayed region despite the very broad lensing redshift kernel.  The values of wavenumber $k$ and redshift $z$ that contribute to a given angular scale $\ell$ in the Limber approximation are shown by the black dotted lines.  The purple dashed line shows the free-streaming scale $k_\mathrm{fs}(z)$ for standard neutrinos with $\sum m_\nu = 58~\mathrm{meV}$, with the region to the right of that line expected to experience suppression due to massive neutrinos. Nonlinear corrections to the matter power spectrum are expected to be non-negligible to the right of the red dash-dot line.  Combined analysis of these probes of clustering should make it possible to disentangle effects that impact the growth of structure on small scales. Figure reproduced from \cite{Green:2021xzn}.}
    \label{fig:Pk_stagger}
\end{figure}

Cosmological probes of the clustering of matter and its suppression due to the presence of massive neutrinos~\cite{Green:2021xzn} and other cosmological relics include gravitational lensing of the CMB~\cite{Kaplinghat:2003bh, Lewis:2006fu}, clustering and weak lensing of galaxies~\cite{Hu:1997mj,Cooray:1999rv}, and the number density of galaxy clusters~\cite{Abazajian:2011dt,Carbone:2011by,Ichiki:2011ue}.
Complementary measurements of cosmological clustering across various scales and redshifts will allow the suppression of clustering due to relics to be disentangled from nonlinear gravitational effects and baryonic feedback that otherwise acts as a source of theoretical uncertainty on the small scale power spectrum.
Figure~\ref{fig:Pk_stagger} shows how various observational probes of cosmological structure are sensitive to the matter power spectrum on different scales and redshifts.
Ongoing and future efforts to measure the cosmological impact of neutrino mass and related effects will enhance the value of small scale structure measurements.

\section{Conclusions}

The advanced new facilities in the upcoming decade, including the next-generation ground-based and space-based GW detectors and telescopes, will enable high data-volume probes of the ultra small-scale structure of the universe for the first time. In this white paper, we laid out the science cases for probing dark matter across a range of observational length scales, from earth-scale to galaxy clusters, but with an emphasis on the new science uniquely probed by the smallest scales. In addition, we discussed the complementarity of future constraints set by measurements of the total matter and radiation content of the universe using next-generation surveys of CMB and the large-scale structure of the universe. 

Coordination of the broad communities, including particle physicists, astronomers, instrumentalists, computer scientists, etc., is required in order to realize the scientific breakthrough using new facilities and data sets. This paper does not only aim to provide perspectives that may guide future decisions about the planning and design of new facilities and surveys on small-scale observations, but also urges the recognition, from the funding agencies and communities, of these new research areas that emerge as new facilities and data become available. None of the novel small-scale probes for dark matter discussed here were included as main science goals when the new facilities were proposed and/or funded, but emerged as the theoretical and observational landscape for dark matter changed. Additional support and recognition, and flexibility in traditional funding streams, are needed to facilitate these cross-field emerging studies in order to achieve the prospects laid out in this paper.

The advances in probing beyond-Standard-Model particles and dark matter through small-scale observations discussed in this white paper will become possible with the data from future observing runs of the current-generation GW detector network and telescopes like the Vera Rubin Observatory and the Nancy Grace Roman Space Telescope, and will further rely on the next-generation facilities, including Cosmic Explorer, Einstein Telescope, LISA, the successor to Gaia, etc. The critical technology and research required for building these future facilities are more comprehensively discussed in Ref.~\cite{snowmass_facility,snowmass_GWfacility}. The strength of this research program relies on its multi-probe nature and the synergies that the different facilities provide, from new GW detectors, to next generation telescopes, pulsar timing arrays and CMB experiments.

In addition, advanced data analysis, signal processing, and computing techniques are crucial to extracting information from the large amount of data and combining multi-messenger channels.
Theoretical and numerical studies, on the other hand, will facilitate the understanding of astrophysical models and the development of optimal analysis techniques.


\section*{Acknowledgments}

R.B. acknowledges financial support provided by FCT -- Funda\c{c}\~{a}o para a Ci\^{e}ncia e a Tecnologia, I.P., under the Scientific Employment Stimulus -- Individual Call -- 2020.00470.CEECIND. 
CD is partially supported by US~Department of Energy~(DOE) grant~\mbox{DE--SC0020223}. 
JGB acknowledges support from the research project  PGC2018-094773-B-C32 [MICINN-FEDER], and the Spanish Research Agency (Agencia Estatal de Investigaci\'on) through the Grant IFT Centro de Excelencia Severo Ochoa No CEX2020-001007-S, funded by MCIN/AEI/10.13039/ 501100011033.
JM is supported by the US~Department of Energy under Grant~\mbox{DE--SC0010129}. 
A.L.M. is a beneficiary of a FSR Incoming Post-doctoral Fellowship. 
L.S. acknowledges the support of the Australian Research Council Centre of Excellence for Gravitational Wave Discovery (OzGrav), Project No. CE170100004. 
Support for S.S. was provided by the Charles E. Kaufman Foundation of the Pittsburgh Foundation.
K. K. Y. N., a member of the LIGO Laboratory, acknowledges the support of the National Science Foundation through the NSF Grant No. PHY--1836814. LIGO was constructed by the California Institute of Technology and Massachusetts Institute of Technology with funding from the National Science Foundation and operates under Cooperative Agreement No. PHY--1764464.
S.C. gratefully acknowledges support from NSF AAG grant 2009828.

\bibliographystyle{unsrt}
\bibliography{main.bib}

\end{document}